# CLASSIFYING CELESTE AS NP COMPLETE

Zeeshan Ahmed[1], Alapan Chaudhuri[1], Kunwar Grover[1], Ashwin Rao[1], Kushagra Garg[1] and Pulak Malhotra[1]

[1]International Institute of Information Technology, Hyderabad
zeeshan.ahmed@research.iiit.ac.in,
alapan.chaudhuri@research.iiit.ac.in,
kunwar.shanjeet@students.iiit.ac.in, ashwin.rao@students.iiit.ac.in,
kushagra.garg@research.iiit.ac.in, pulak.malhotra@students.iiit.ac.in

## ABSTRACT

*We analyze the computational complexity of the video game "CELESTE" and prove that solving a generalized level in it is NP-Complete. Further, we also show how, upon introducing a small change in the game mechanics (adding a new game entity), we can make it PSPACE-complete.*

## 1. ABOUT CELESTE

CELESTE [1] is a single-player 2D platformer developed by Maddy Thorson and Noel Berry. The game is about you being Madeline who is climbing to the peak of the mountain "Celeste". The goal of the game is to overcome obstacles on the way and make it to the end of the level. CELESTE consists of 7 levels which consist of sub-parts. In each subpart, you have to reach an exit point without taking damage. Taking damage resets you back to the starting point.

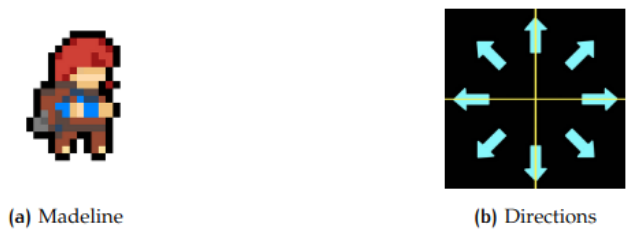

Figure 1. CELESTE player

### 1.1. About the Player

Madeline is the main character whose actions are governed by our controls. She is restricted in 8 directions of motion and primarily has 3 special moves other than basic left and right movement. Those are:

**Directions of movement:** Madeline can move in 8 directions as described in the left diagram. These and the other moves have fixed button/button combinations assigned to them.

**Jump:** She has the ability to jump to a certain height, which can be then done again only when she hits the ground.

**Dash:** When Madeline has the "charge" she can dash in any direction, this gives her extra momentum and the ability to dash into "space blocks" (described later). The charge gets used up when she dashes and is restored when she hits the ground or passes through the space block. There are other objects which also recharge her, but those are not used in the proof.

There are other objects which also recharge her, but those are not used in the proof.

**Grab:** Since she is a mountain climber, she has the ability to climb walls and wedges. But she has stamina, which limits the amount of time she can grab onto a wall before sliding down. This stamina is restored when she makes contact with the ground.

## 2. LEVEL IMPLEMENTATION

### 2.1. Frames

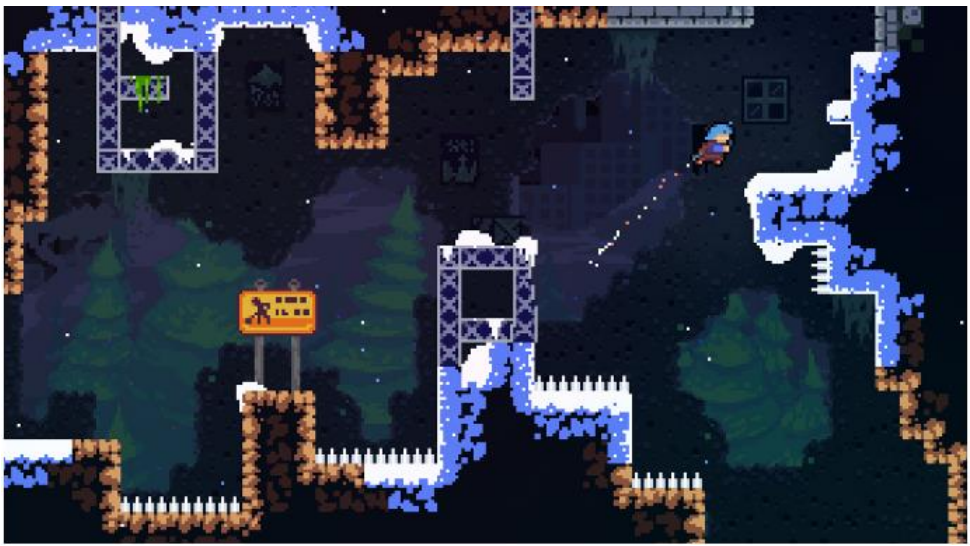

Figure 2. Bottom left corner is the entry point and the top right corner is the exit point

Frames are areas of games which has a start and an endpoint. You have access to one frame at a time. Using the entries and the exits, you move from 1 frame to another. So Frames serve as the checkpoints in the game. Frames do not have a limit of size since your screen can scroll.

The game has been split into such frames, which are puzzles on their own, which require planning and reflexes to reach the endpoint. There are multiple ways to solve these puzzles due to the game's versatility and the mechanisms in it.

A graph of such frames connected together makes up a level of the game. To construct our proof, we will make frames which we will join together to make our level.

## 2.2. Objects

Revolving around the basic 3 operations the game has a lot of mechanisms that add to the fun and the difficulty of the game. These are added to the game as the player makes progress in the levels.

For our proof, we will mainly use 3 of the objects. They are:
- Unstable Platform
- Button door
- Space block

We explain how these objects work below. The purpose of these objects will be explained later.

### 2.2.1. Unstable Platform

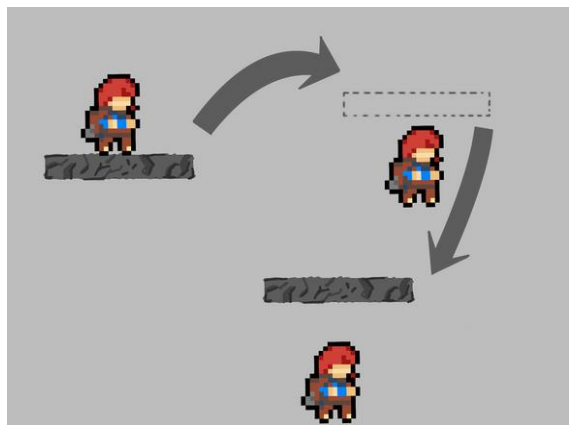

Figure 3. Unstable platform

A stone platform that can float, this platform breaks when Madeline stands on this platform for more than a second. After the platform breaks, Madeline if not jumped will fall down. This platform then forms back in the same place. But it can only be broken when stood upon and can not be broken from below, making it like a trap door if placed correctly.

### 2.2.2. Button Door

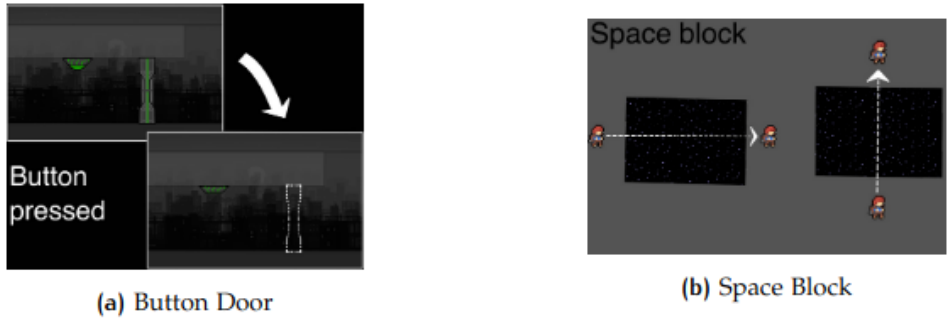

Figure 4. Button door and space block

A door that can only be unlocked using its specific button. This button must be placed in the same frame as the door, but it has the freedom to be placed anywhere in the frame. Once the door is opened it cannot be closed again.

**2.2.3. Space Block**

A block of celestial material which lets you float into and out in a straight line when you dash into it. Once Madeline dashes into the block, you cannot stop her from reaching the other opening of the block in a straight line. If the other side of the line is blocked with a wall, Madeline dies and respawns at the start of the frame.

**2.3. Levels**

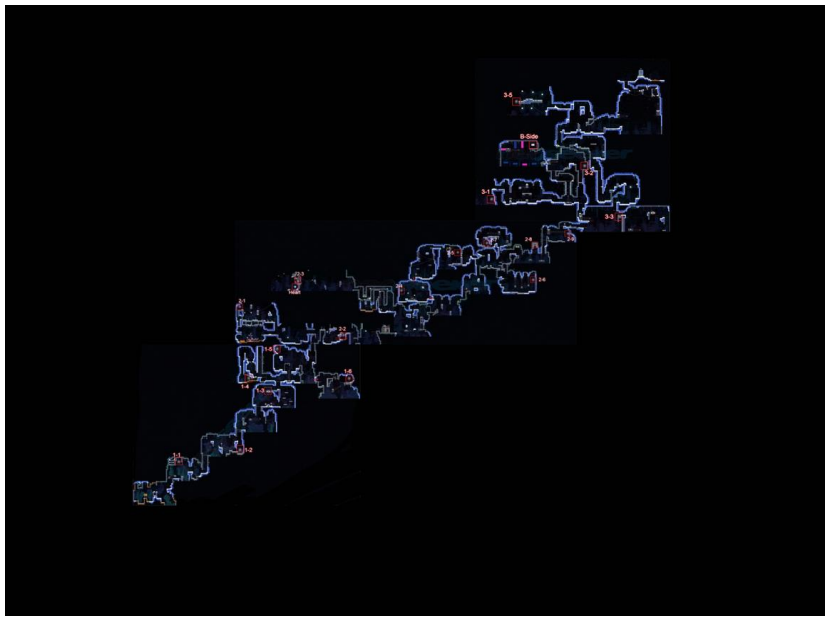

Figure 5. Bottom left being the start and the top right being the end

A level consists of many frames, but there is only 1 flag at the top of the level and there is only 1 initial start point of the level. For most of the levels, there is only 1 linear path from the start to end with a sequence of frames, but some other levels are more complex, which include sub-tasks and other detours. Here is an example, this is the 1st level in the game, the frames have been arranged according to the order.

## 3. COMPLEXITY CLASSIFICATION

Now that the game has been well defined, we work on classification of the game. Whenever we say "CELESTE belongs to *X* complexity class", we mean to say that the decision problem of deciding whether finish point is reachable from start point.

### 3.1. Basic Observation

Given a level and a path, that is the moves required to reach the endpoint, you can verify if the path is correct just by applying those moves. The moves are polynomial, why? We repeat the sections only after visiting other sections. Since the number of sections themselves are polynomial, we can only have polynomial moves before we complete the level.

This clearly implies that the game is NP [2]. For example, for the 1st level, we can map the path from start to end as seen below.

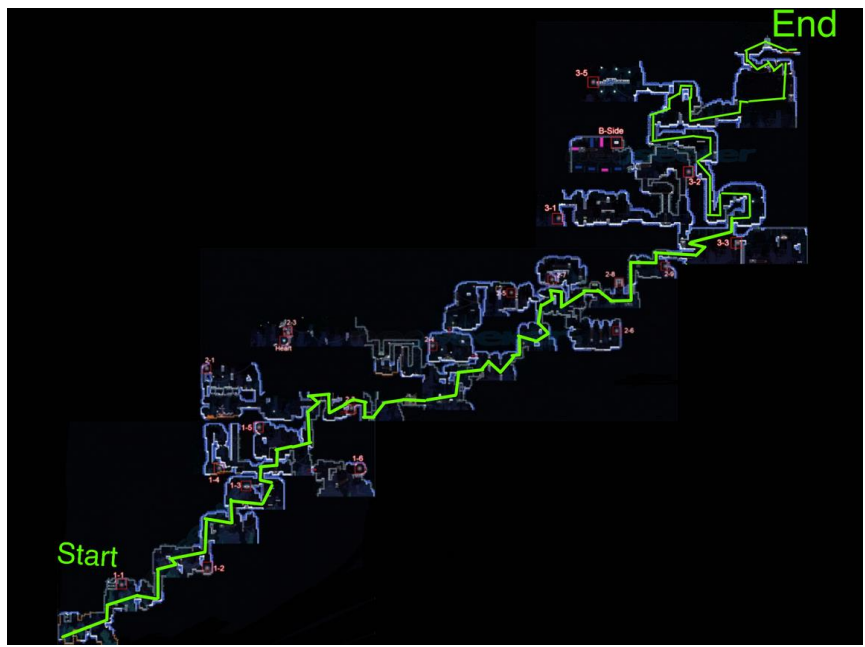

Figure 6. In this pattern we can always map from start to end

Now since we have proven that CELESTE is NP. We can try to prove that it is NP-Hard [2], essentially proving that it is NP-Complete [2].

### 3.1.1. Intuition behind being NP-Hard

Due to many paths of the game which do not lead to the end and certain mechanisms that lock us out which we will see in the future, it is not possible to tell whether there exists a polynomial-time algorithm to solve the levels. So, we will rather try proving that this game is NP Hard.

## 4. FRAMEWORK FOR NP-HARDNESS

To prove the NP-hardness of CELESTE, we here describe a framework for reducing 3-SAT [2] to a 2-D platform game. This framework is based on (link source here). Using this framework in hand, we can prove the hardness of games by just constructing the necessary gadgets [3-5].

The framework is a reduction of the 3-SAT problem. We start from the Start gadget, and end at the Finish Gadget. At each Variable gadget, we make a choice and turn on the Clause gadgets according to the choice, which in essence is making a literal true or false.

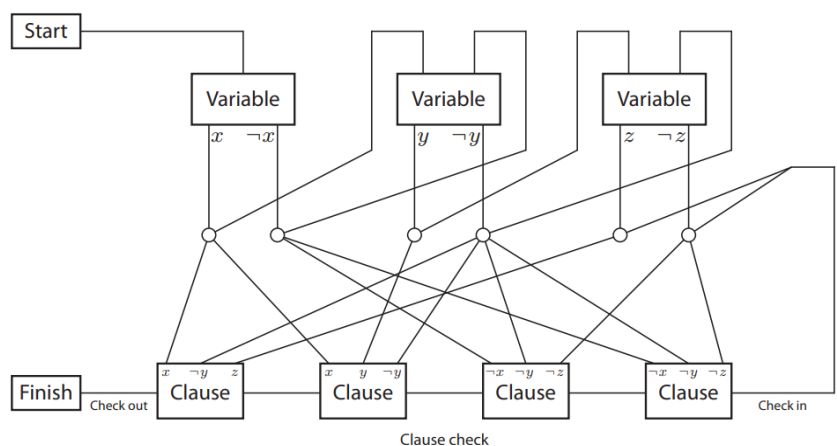

Figure 7. General framework for NP-Hardness

At the end, we make a pass through all of the clauses sequentially and we are able to pass through them if all of them are satisfied. Since the game is 2D, we also need a Crossover gadget. The Crossover gadget makes sure that if there are two overlapping connections, we travel the connections one by one. Below given are more details for the gadgets.
- **Start and Finish:** The start and end gadgets contain the spawn point and the end goal respectively.
- **Variable:** Each variable gadget must force the player to make a binary choice (select $x$ or $\sim x$). Once a choice is taken the other choice should not be accessible. Each variable gadget should be accessible from and only from the previous variable gadget is such a way that it is independent of the choice of the previous gadget and going back is not allowed.
- **Clause:** Each literal in the clause must be connected to the corresponding variable. Also, when the player visits the clause, there should be a way to unlock the corresponding clause.
- **Check:** After all the variables are passed through, all the clauses are run through sequentially. If the clause is unlocked, then the player moves on to the next clause else loses.

- **Crossover:** The crossover gadget allows passage via two pathways that cross each other. The passage must be such that there is no leakage among them.

If we can build these gadgets using a game, we can reduce 3-SAT to that game using this framework and show that the game is NP-Hard.

## 5. CELESTE IS NP-HARD

To prove that CELESTE is NP-hard, we will try to use the above framework to reduce 3-SAT to CELESTE.

### 5.1. Variable Gadget

A Boolean Variable can take 2 values, True or False, and it might have multiple occurrences throughout the formula.

The verification of 3-SAT is done by giving the satisfiable values to the variables, hence the values cannot be changed in the middle of the substitution. For now, we need to take care of the binary and the irreversible nature of boolean variables. We do this with the help of an Unstable Platform.

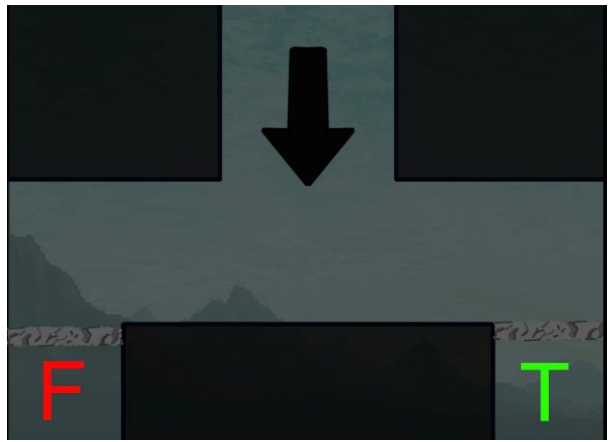

Figure 8. Exits are covered by Unstable platforms making them one way traps

These exits have an Unstable platform covering them, these have to be broken before the exit can be used. But, why the unstable platform?

Madeline falls from the top on the platform, that is the only entry to the Gadget. The Gadget has 2 exits on the sides of the floor, each leading to a tunnel. The unstable platform makes Madeline seal her choice. Once the path is taken, there is no way to access this frame again other than restarting since the platform will reform blocking the entry.

### 5.2. Clause Gadget

Each Clause has 3 variables, out of which even if 1 were true the Clause would be true. To implement that in our frame, we use The Button Door.

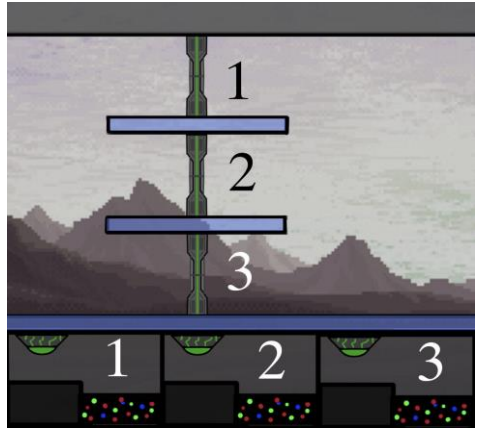

Figure 9. The colourful dots represent space block and the hits are reachable by Madeline

A Clause gadget consists of 3 Parallel Button Doors, the buttons are accessed through the variable tunnels. For now, do not worry about how the tunnels are connected. The main idea is that even if 1 door opens it is sufficient for Madeline to pass through the region.

Madeline presses the buttons according to the values she took for the variables, these will unlock the doors, if the variables made a clause true, the clause would have at least 1 door open.

## 6. SUPPORT GADGETS

Now these above-mentioned gadgets must be connected and for that, we use our support gadgets that will be constructed as per the requirements.

### 6.1. The Tunnel

To connect the Variable exit to the Buttons of the Clause, we use a Tunnel gadget.

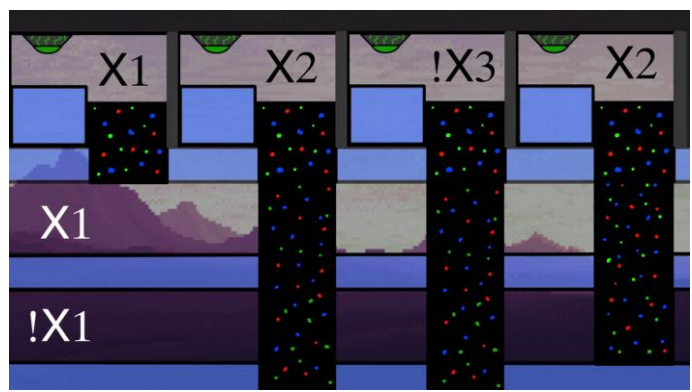

Figure 10. $x_1$ Tunnel only has access to button which have $x_1$ as their variable in their clause

Below the buttons, there are Tunnels leading from the exits of the variables. The variables have access to the buttons they can set true according to the Boolean expression.

For example, if $x_1$ is chosen to be true, then Madeline gets access to the $x_1$ tunnel and $\sim x_1$ if she had chosen false. $x_1$ tunnel has access to buttons that open a door to the clause having $x_1$.

How do we block the variables from accessing the other doors? For that, we have constructed the Crossover Gadget. The space blocks that are displayed in the diagram are used in a specific manner described in the Crossing Frame.

### 6.1. Crossover Gadget

Since the game is 2D, you cannot avoid paths from crossing each other during the construction of such a level. We can make sure that the intersection of the paths happens only in the form of a cross.

Suppose we want Madeline to go from $A_1$ to $A_2$ or $B_1$ to $B_2$ or the other direction. But she shouldn't be able to go from an *A* to a *B* or vice versa.

The Space block as described before teleports the player from one end to the other in a straight line without any interference. Encountering a wall will kill Madeline and she will respawn at the start of the frame.

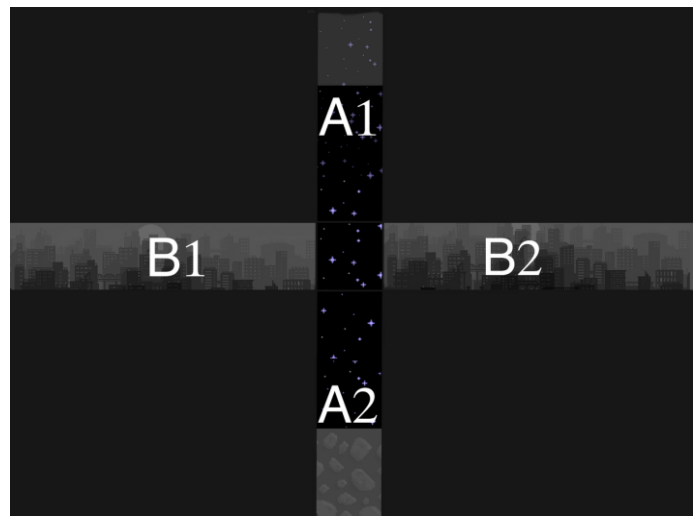

Figure 11. Crossover gadget

So, we put the space blocks in the intersection of the paths in such a way that there are no straight lines connecting the opening side of *A* to *B*. It might seem like you can draw a line from *A* to *B* but remember that Madeline can only move in 8 directions, so the lines can be parallel or 45 degrees inclined with the axis. So, no such line will exist. This means that the only way she can travel through the space block is in a straight line parallel to the axis, hence she cannot access *A* from *B* or vice versa.

# 7. SEQUENCE OF FRAMES

Now that all the frames have been constructed, we decide the sequence of the frames. Let $n$ be the number of variables in the Boolean expression distributed into $k$ clauses.

## 7.1. Variable Order

We will assume the order of the variables to be $x_1, x_2, x_3 \ldots x_n$. We select values for these variables in the same order. So the starting position will be in the $x_1$ gadget since we pick its value first.

## 7.2. Transition Between Variables

From the variable gadget of $x_1$, we choose its value and go to the respective tunnel. In the tunnel, we press all the accessible buttons, after all the buttons, we reach the end of the tunnel. Now since the values of $x_1$ have been already picked and substituted, we have to pick a value for the next variable hence we must go to the $x_2$ variable gadget.

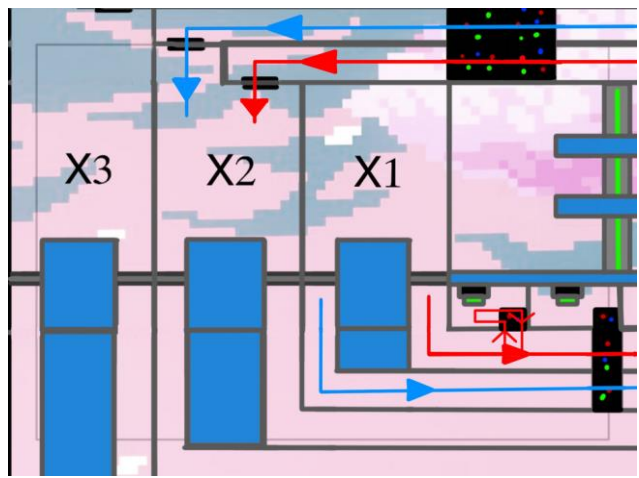

Figure 12. Transition from $x_1$ to $x_2$ after completing the path

In such order we pick the value of all the variables, click the buttons which open the clause doors, and continue until we reach the end of the $x_n$ variable. Since we ran out of variables, where do we go now?

## 7.3. Final Passage

At the end of the $x_n$ passage, we should have an entrance to the Final passage containing the clauses. Madeline after choosing all the variables will now have to reach the flag.

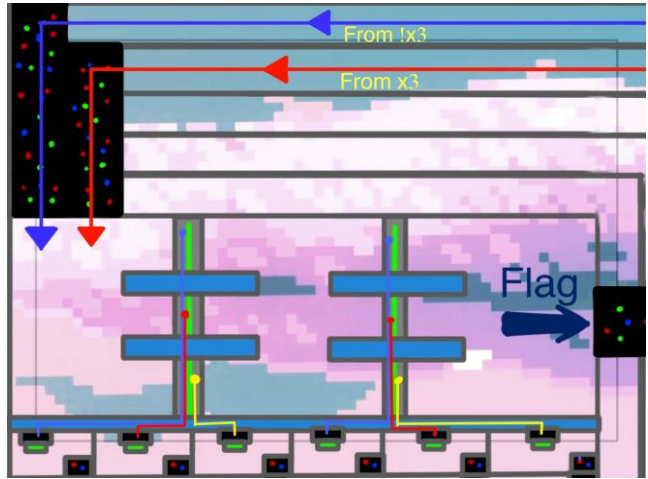

Figure 13. Example case where n = 3

The path to the flag lies on the other side of the passage. If at least one door from each Clause is open only then will she be able to reach to the flag, else she won't be able to complete the level. This was the AND of all the OR clauses, leading to a normal form with 3 variables in each clause.

### 7.3.1. Why not put the flag at the end?

The flag is always at the top of the level. So we need to redirect the player from the end of the final passage to the flag. Since there are other Paths that come between it, we use crossover frame.

## 8. FINAL LEVEL

Now that all the separate parts have been explained, we put together our final level. The Boolean expression for which the level has been implemented is:

$$(x_1 \vee x_2 \vee \neg x_3) \wedge (\neg x_1 \vee x_2 \vee x_3)$$

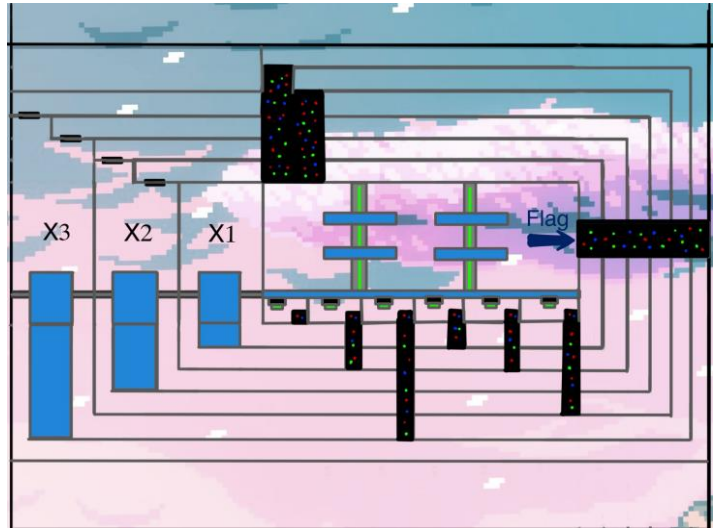

Figure 14. Final level layout

## 9. EXAMPLE SUBSTITUTION

Let the substitution of the variables be:
- $x_1 = 1$
- $x_2 = 0$
- $x_3 = 1$

We start with the $x_1$ gadget, take the true tunnel, and activate the door. After which we end up with a tunnel with an exit leading to $x_2$ gadget.

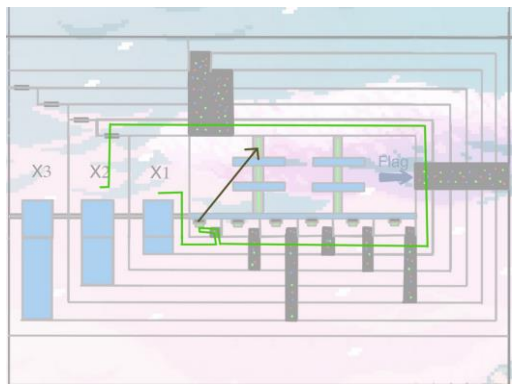

Figure 15. Going from x1 to x2 gadget

Now we are in the $x_2$, we take the false tunnel, but since there is no $x_2$ in the expression, there is no door that can be opened from this tunnel. So, we just continue and end up in the $x_3$ frame. In the $x_3$ frame, we repeat the same procedure. We open a door in the 2nd OR clause and end up at the end of the tunnel which leads to a space block. This space block when used will take Madeline to the beginning of the 'final passage.'

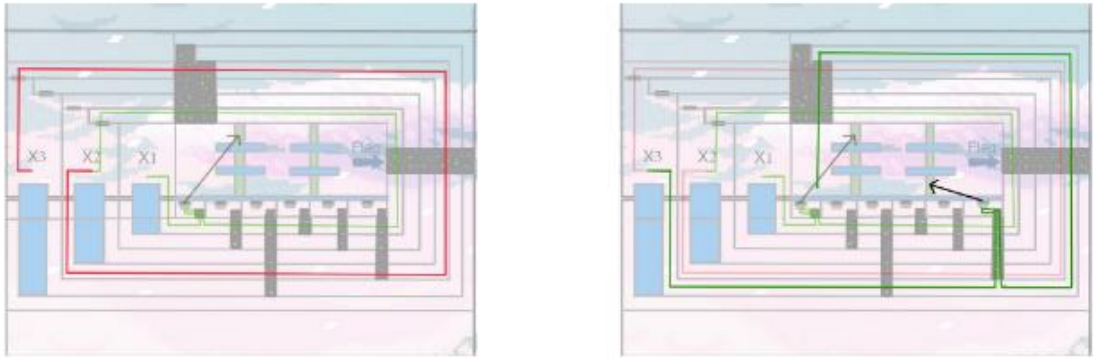

Figure 16. Going to the Final passage

Now one door of each clause has opened, Madeline can pass the final passage and go to the flag.

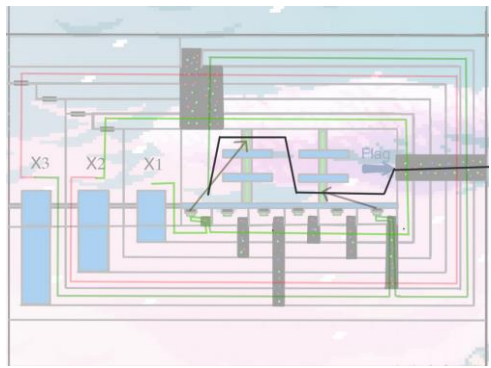

Figure 17. Reaching the flag

Since Madeline was able to reach the flag. The level could be completed, hence the expression as expected is satisfied with the given values.

## 10. SUMMARY ON NP-COMPLETENESS

This proves that CELESTE is at least as hard as 3-SAT, making it NP-Hard. In conclusion, the game is both NP and NP-Hard, making it an NP-Complete puzzle.

## 11. MAKING CELESTE PSPACE-COMPLETE

We now describe how making a small change to CELESTE can make it PSPACE-Complete [2]. In the proof the CELESTE is NP-Complete, we used a button door which opened when we pressed the green button and then was obsolete.

We make an addition to the game. The door can now be closed using a red button. When the door is open, the red button is deflated and can be activated and when the door is closed the green button can be activated.

## 12. CELESTE BELONGS TO PSPACE

To prove that CELESTE is a PSPACE-Complete puzzle, we have to first show that it belongs to PSPACE [2]. Now, it is sufficient to show that CELESTE belongs to NPSPACE [2] since NPSPACE is a subset of PSPACE by Savitch's Theorem [6]. This means that for any given traversal on the level, it has to use polynomial space with respect to the size of the level. Since, the game's element behaviour is a simple (deterministic) function of the player's moves. Therefore, we can solve a level by making moves non-deterministically while maintaining the current game state (which is polynomial).

## 13. FRAMEWORK FOR PSPACE-HARDNESS

To prove PSPACE-Hardness of CELESTE, we here describe a framework for reducing TQBF problem [2] to a 2-D platform game. This framework is based on (link source here). Using this framework in hand, we can prove hardness of games by just constructing the necessary gadgets. For this framework we need one more gadget: **Pressure Button Door Gadget.**

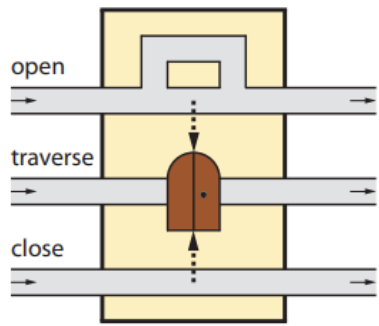

Figure 18. Pressure Button Door Gadget

In CELESTE, to press the button, Madeline has to dash into it. The button in the above frame is thick enough to prevent Madeline to pass through the tunnel without pressing the button. Thus, it forces her to press the button.
- The open path has a button which the player is forced to press.
- The traverse path is the path containing the door which can be traversed if the door is opened.
- The close path button forces the player to close the door.

### 13.1. A General Framework for PSPACE-Hardness

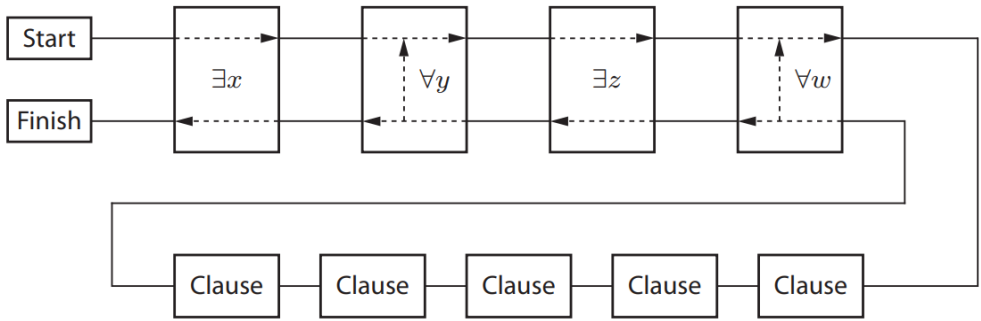

Figure 19. Framework

A given fully quantified Boolean formula

$$\exists x \forall y \exists z \ldots \phi(x, y, z, \ldots)$$

where PHI is in 3-CNF is translated into a row of Quantifier gadgets, followed by a row of Clause gadgets, connected by several paths.

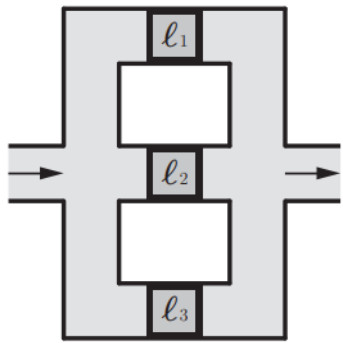

Figure 20. Clause gadget

The clause gadget is built using three gates whose pressure buttons are in quantifier gadgets. And, for building Quantifier gadgets we use a special notation in a tunnel as shown below:

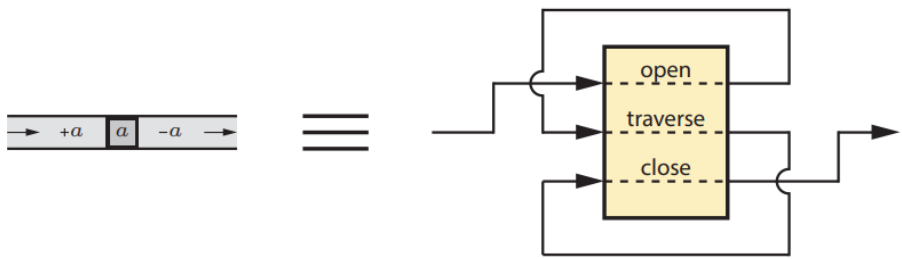

Figure 21. Shorthand notation for tunnels

Here, $+a$ opens the gate corresponding to variable and $a$ and $-a$ closes the gate corresponding to gate $a$. The player is forced to press these buttons as described before.

## 13.2. Existential Quantifier

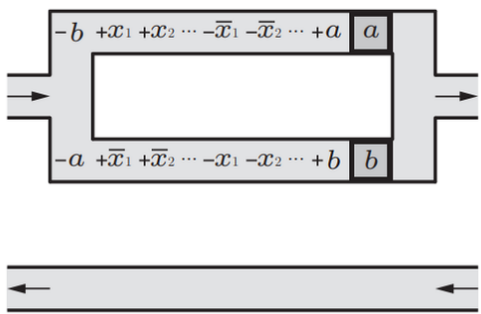

Figure 22. Shorthand notation for tunnels

The player can only select one of the path ways and once it is selected, the player can never change his choice.

## 13.3. Universal Quantifier

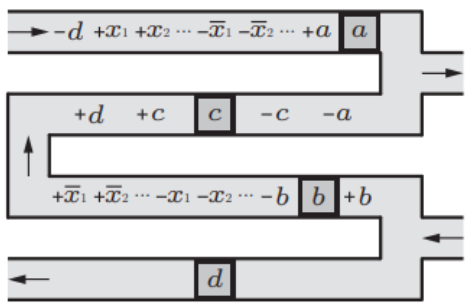

Figure 23. Shorthand notation for tunnels

The player first proceeds to mark the variable as true and while backtracking has to traverse through the clauses again with the variable marked as false. After both possibilities are tried the player is able to move forward.

## 13.3. Conditions for Traversal

Traversing a quantifier gadget sets the corresponding variable in the clauses. When passing through an existential quantifier gadget, the player can set it according to their choice. For the universal quantifier gadget, the variable is first set to true.

A clause can only be traversed if at least one of the variables is set in it. After traversing all the quantifier gadgets, the player does a clause check and is only able to pass if all the clauses are satisfied. If the player succeeds, they are routed to lower parts of the quantifier gadgets, where they are rerouted to the last universal quantifier in the sequence.

The corresponding variable is then set to False and the clauses are traversed again. This process continues and the player keeps backtracking and trying out all possibilities. We will later show how to build these quantifier gadgets using CELESTE game entities.

## 14. MODIFIED CELESTE IS PSPACE-HARD

Using the previously described framework, we will build corresponding gadgets and thus show a reduction from the TQBF problem to CELESTE, implying the CELESTE is PSPACE-Hard.

### 14.1. Door Gadget

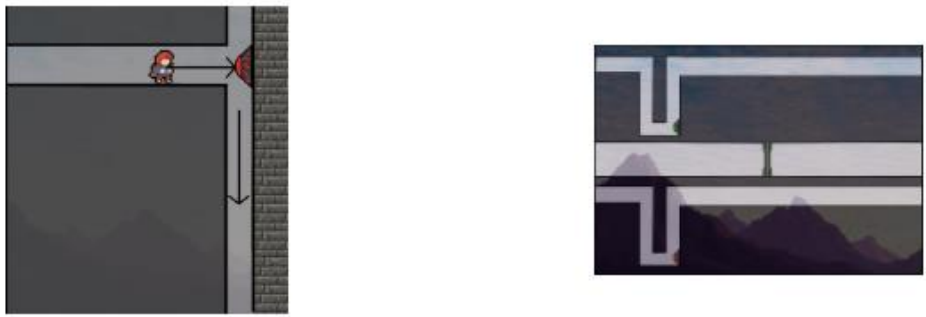

Figure 24. Door gadget and force button

For creating a door gadget we first need to create a way to force the player to dash into a button to activate it. We do this using a narrow tunnel and leaving only enough space to pass if the button is pressed. Now, using this force button, we create a door gadget. The button above will open the door, and the button below will close the door, and since the path is thin, Madeline will not be able to pass through until the button is pressed.

### 14.2. Multi-Tunnel Gadget

For clubbing multiple open/close symbols together, we use a multi tunnel gadget. This is just a helpful gadget to make Quantifier Gadgets.

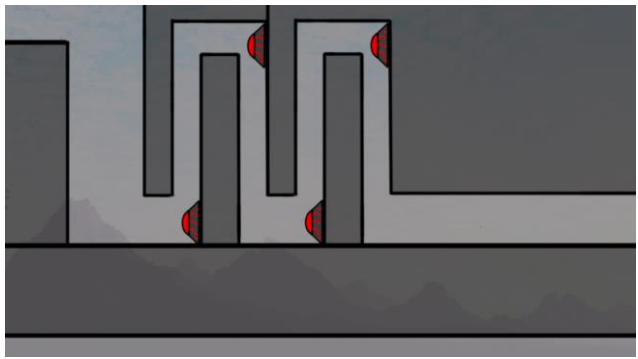

Figure 25. Multi-tunnel gadget

## 14.3. Putting it all together

Since we were able to make the gadgets described in the framework, we have a reduction from the TQBF problem to Modified CELESTE. Thus, Modified CELESTE is PSPACE-Hard.

## 15. SUMMARY ON PSPACE-COMPLETENESS

In conclusion, the modified game is both PSPACE and PSPACE-Hard, making it a PSPACE-Complete puzzle.

On adding the close button, we added a requirement to keep track of all the doors that are open. Before once a door was open, it always remained open. Before we had to only keep track of the current state sequentially as all the doors will be opened in a sequence. Knowing that a door is openly implied that all the previous doors were opened to reach the current door.

But after adding the close button, at any point of the game, all the doors are independent and knowledge of the open state of a door gives us no info about the other doors. So, we must keep track of all the other doors. This makes the game harder and makes its PSPACE-Hard instead of NP-Hard.

In short, lack of knowledge about the status of all the doors makes the game PSPACE-Complete.

## 16. CONCLUSIONS AND FUTURE WORK

In this paper, we have proven that the task of solving levels for the original version of the video game CELESTE is NP-Complete. As described above, this means that this problem is equivalent to solving the Boolean satisfiability problem. Furthermore, addition of complications to the original version in the form of a door-closing button results in the modified game becoming harder [7], PSPACE-Complete to be specific.

This work serves as a stepping stone towards analysing computational complexity of more sophisticated interactive systems and increases the scope of video games that have been studied within this field. Further work involves investigating the hardness of platformers with more involved mechanics, similar to what has been done on analysing the complexity of the popular arcade game Angry Birds [8]. We are hopeful that our work opens up new directions to understand the relationship between the mechanics of interactive systems and computational complexity.

**Authors**

Zeeshan Ahmed

Senior, Bachelor of Technology and Master of Science Dual Degree in Computer Science at International Institute of Information Technology, Hyderabad.

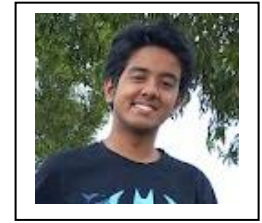

Alapan Chaudhuri

Senior, Bachelor of Technology and Master of Science Dual Degree in Computer Science at International Institute of Information Technology, Hyderabad.

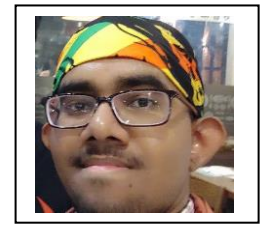

Kunwar Grover

Senior, Bachelor of Technology in Computer Science and Engineering at International Institute of Information Technology, Hyderabad.

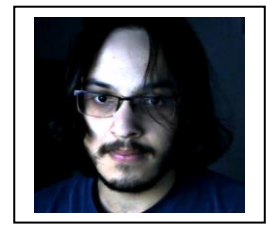

Ashwin Rao

Senior, Bachelor of Technology in Computer Science and Engineering at International Institute of Information Technology, Hyderabad.

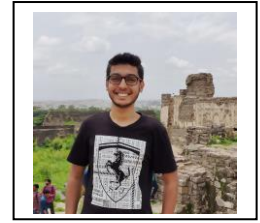

Kushagra Garg

Senior, Bachelor of Technology and Master of Science Dual Degree in Computer Science at International Institute of Information Technology, Hyderabad.

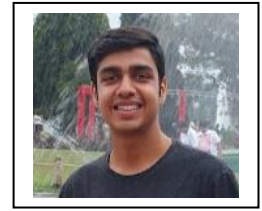

Pulak Malhotra

Senior, Bachelor of Technology in Computer Science and Engineering at International Institute of Information Technology, Hyderabad.

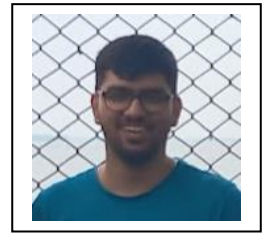